# Party Autonomy in Determining the Law Applicable to Non-contractual Obligations concerning Cross-Border Data Transfers

Yuki Okamura[1]*, Ren Yatsunami[2], Kumiko Kameishi[1], Oliver Posani[3], Soma Araoka[1], Miho Ikeda[1], Makiko Aoyagi[1]

[1] NTT Social Informatics Laboratories, Tokyo, Japan
[2] Faculty of Law, Kyushu University, Fukuoka, Japan
[3] EY Law, Vienna, Austria
* Corresponding author: Yuki Okamura
   Email: yu.okamura@ntt.com

**Key Points**

- Cross-border data transfers have become a matter of daily occurrence against the backdrop of the development of cloud computing and artificial intelligence. Consequently, where a data leak gives rise to civil liability, the determination of that liability inevitably assumes an international dimension involving foreign elements.
- As is starkly demonstrated by secret sharing technology in cloud computing, fragments of data may be presumed to be distributed across multiple jurisdictions on a global scale. This renders traditional private international law measures — predicated on the identification of a physical location — inadequate for the purposes of determining the applicable law, a difficulty that is particularly acute in relation to non-contractual obligations.
- Bearing in mind the typical scenario encountered in practice — in which a Data Subject brings a claim for damages against a SaaS (Software as a Service) provider, which in turn seeks recourse against an IaaS (Infrastructure as a Service) or PaaS (Platform as a Service) provider — a characteristic feature of such cases is the concurrence of contractual and non-contractual obligations. Taking this feature into account, it is possible to determine the applicable law governing non-contractual obligations through party autonomy — by aligning it with the law governing the contractual obligation as selected by the parties, an approach that may be termed private ordering. This serves to overcome the difficulties associated with the identification of a physical location and, at the same time, contributes to ensuring the foreseeability of the parties.

## 1. Introduction

  This paper examines, through a comparative legal study of EU law and Japanese law, whether party autonomy can influence the choice of law applicable to non-contractual obligations, with particular reference to cross-border data transfer.

  Two technological developments have particularly accelerated cross-border data transfers in recent years.

  First, in the development of AI—large language models (LLMs) in particular—it has become evident that model performance improves as the volume of training data (strictly speaking, the number of tokens) increases[1]. Moreover, it is widely observed that training on data in one language can contribute to improved capabilities in other languages—a phenomenon known as cross-lingual generalization[2]. In response, vast quantities of training data are now being acquired through cross-border data transfers in the pursuit of building better-performing models.

  Second, cloud-based services—deployed on cloud infrastructure such as AWS and Azure—have emerged as an alternative to on-premises systems[3], in which infrastructure is hosted on physical servers maintained locally[4]. In deploying such cloud infrastructure, systems are increasingly being built and services delivered using data centers located in other countries, driven by considerations of technical and business efficiency as well as economic security or business continuity.

  These developments have rendered cross-border data transfers ubiquitous, thereby making the question of the applicable law for non-contractual liability arising in connection with such transfers all the more pressing.

  With regard to cross-border data transfers as described above, extensive research has focused on public-law aspects of data protection law, including the relationship between regulatory authorities and business operators, particularly administrative fines such as those under Article 83 of the GDPR[5][6]. Such research has addressed, in particular, regulations governing the provision of personal data to third parties in foreign jurisdictions such as Articles 44 to 49, and the corresponding administrative penalties under Article 83(5)(c) of the GDPR, as well as to the extraterritorial application of data protection laws such as Article 3 of the GDPR[7].

---

[1] Jared Kaplan and others, 'Scaling Laws for Neural Language Models' (arXiv, 23 January 2020) <https://arxiv.org/abs/2001.08361v1> accessed 30 March 2026.

[2] Niklas Muennighoff and others, 'Crosslingual Generalization through Multitask Finetuning' in Proceedings of the 61st Annual Meeting of the ACL (Volume 1: Long Papers) (Association for Computational Linguistics 2023) 15991; Alexis Conneau and others, 'Unsupervised Cross-lingual Representation Learning at Scale' in Proceedings of the 58th Annual Meeting of the Association for Computational Linguistics (Association for Computational Linguistics 2020) 8440.

[3] Michael Armbrust and others, 'A View of Cloud Computing' (2010) 53(4) Communications of the ACM 50.

[4] The technical definition of Cloud Computing can be found in Peter Mell and Timothy Grance, 'The NIST Definition of Cloud Computing' (NIST Special Publication 800-145, September 2011) <https://doi.org/10.6028/NIST.SP.800-145> accessed 5 March 2026.

By contrast, private-law aspects—concerning, for instance, the relationships between data subjects and service providers, or between service providers and cloud computing operators—in cases where data breaches and similar incidents give rise to civil liability have remained largely underexplored. Where such liability involves cross-border elements, the determination of the applicable law becomes a question of private international law. Such civil liability may be contractual or non-contractual in nature. With respect to contractual obligations, party autonomy has become the prevailing global norm, allowing the parties to choose the applicable law, as provided for in Article 3 of the Rome I Regulation[8] and Article 7 of the Act on General Rules for Application of Laws in Japan (hereinafter "AGRALJ")[9]. By contrast, the position regarding non-contractual obligations varies considerably from one jurisdiction to another.

As this paper employs a comparative law approach between EU law and Japanese law, as discussed in greater detail below, the following section illustrates, by reference to the legal frameworks of these two jurisdictions, that the permissible scope of party autonomy with respect to non-contractual obligations differs between them.

Under Japanese private international law, the parties to a tort may agree on the governing law of tort only after the tort occurs (Article 21 of the AGRALJ). However, it would not be accurate to understand that the significance of prior agreement on the governing law in this area is completely denied. Rather, it can be observed that the existence of an exceptional clause (Article 20 of the AGRALJ) creates a framework in which the autonomy of the parties in the governing law of a contract spreads to the area of torts.

---

[5] Regulation (EU) 2016/679 of the European Parliament and of the Council of 27 April 2016 on the protection of natural persons with regard to the processing of personal data and on the free movement of such data, and repealing Directive 95/46/EC (General Data Protection Regulation), OJ 2016 L 119/1.

[6] For example, Waltraut Kotschy, 'Article 83: General Conditions for Imposing Administrative Fines' in Christopher Kuner and others (eds), The EU General Data Protection Regulation (GDPR) (OUP 2020) 1180.

[7] Benjamin Greze, 'The Extra-Territorial Enforcement of the GDPR: A Genuine Issue and the Quest for Alternatives' (2019) 9(2) International Data Privacy Law 109.

[8] Regulation (EC) 593/2008 of the European Parliament and of the Council of 17 June 2008 on the law applicable to contractual obligations (Rome I), OJ 2008 L 177/6.

[9] For an English translation, see Japanese Law Translation Database System <https://www.japaneselawtranslation.go.jp/ja/laws/view/3783> accessed 30 March 2026.

Turning to the European situation on private international law, the Rome II Regulation[10] stipulates several provisions that take into account the parties' intentions.

How (positively or negatively) should we evaluate this modern trend in the choice-of-law framework, which has come to give the parties' intentions a certain influence on courts' decisions on the applicable law in tort cases?  In other words, this is the question of the extent to which the party autonomy principle should be respected in choice-of-law issues concerning non-contractual obligations. To address this question, this paper focuses on regulations governing cross-border data transfers mainly for two reasons outlined below.

First, with the development of influential data protection legislation such as the GDPR, an increasing role for standard clauses, such as privacy policies, in the governance of cross-border data transfers has been observed, which is linked to the discussion of the relationship between the governing law of contract and that of tort.

In particular, empirical studies have demonstrated that, following the enactment of data protection legislation, privacy policies are drafted or amended in a manner that traces the content of such legislation[11]. Where a data leak or similar incident occurs, private-law claims by data subjects—such as claims for compensation or the erasure of personal data—may be grounded in two concurrent legal bases: non-contractual liability under data protection law such as the GDPR or the AGRALJ, and contractual liability arising from the breach of terms shaped by the privacy policy, incorporated through standard terms regulation or comprehensive consent.

In legal terms, this constitutes a concurrence of contractual and non-contractual claims (Konkurrenz von vertraglichen und außervertraglichen Ansprüchen). Where the substantive content of the claims is essentially identical, applying different choice-of-law rules merely on account of the difference in legal basis risks undermining legal certainty, thereby giving rise to a need to treat both claims in a unified manner. Specifically, given the practical

---

[10] Regulation (EC) 864/2007 of the European Parliament and of the Council of 11 July 2007 on the law applicable to non-contractual obligations (Rome II), OJ 2007 L 199/40.

[11] Kevin E Davis and Florencia Marotta-Wurgler, 'Contracting for Personal Data' (2019) 94(4) New York University Law Review; Michael Kretschmer and others, 'Cookie Banners and Privacy Policies: Measuring the Impact of the GDPR on the Web' (2021) 15(4) ACM Transactions on the Web art 20, 1; Henry Hosseini and others, 'A Bilingual Longitudinal Analysis of Privacy Policies Measuring the Impacts of the GDPR and the CCPA/CPRA' (2024) 2024(2) Proceedings on Privacy Enhancing Technologies 434.

difficulty—discussed in greater detail below—of determining the applicable law in tort, it is appropriate to consider an approach that aligns the law applicable to non-contractual obligations with the law applicable to contractual obligations. In the course of this inquiry, the question arises as to the extent to which party autonomy should be taken into account in determining the law applicable to non-contractual obligations.

Second, in the case of cloud-based services that employ distributed architectures, data may be replicated or sharded across nodes located in multiple jurisdictions. In such systems, where a personal data breach occurs—for instance, through hacking or unauthorized access—the breach may involve multiple locations simultaneously and may traverse several nodes, such that no single location can be identified as the place where the breach occurred. As a result, the fundamental approach of private international law—selecting the applicable law by reference to the place where the harmful event occurred—cannot function effectively in such circumstances, necessitating an alternative method of determining the applicable law. Here, there is an expectation that increasing the number of situations in which the applicable law governs contractual agreements—rather than by reference to a physical location—will contribute to greater predictability.

For the purposes of the analysis described above, it is appropriate to use EU law and Japanese law as comparative materials. Both legal systems share the fundamental position of recognizing party autonomy, yet each presents distinctive features that make comparison fruitful. As regards Japanese law, as discussed above, it has introduced a provision that allows the law applicable to non-contractual obligations to be aligned with the law applicable to contractual obligations (Article 20 of the AGRALJ), making it possible to draw insights directly relevant to the foregoing discussion. As regards EU law, it possesses the GDPR, which is widely regarded as a leading framework for data protection legislation with significant global influence[12]. Moreover, the Rome II Regulation permits the choice of the applicable law by prior agreement where all the parties are pursuing a commercial activity (Article 14(1)(b)), and recognizes that a pre-existing contractual relationship between the parties may influence the determination of the law of the country with which the tort is most closely connected (Article 4(3)). In this way, EU law takes a different approach from Japanese law in incorporating party autonomy into the choice of law for non-contractual obligations.

In light of the practical significance of these legal frameworks and the value of comparing the diversity of their approaches, this paper seeks to elucidate the function of party autonomy in the determination of the law applicable to non-contractual obligations through a comparative analysis of EU law and Japanese law.

---

[12] On the global regulatory influence of the GDPR described captured by the term "Brussels Effect", see Anu Bradford, The Brussels Effect: How the European Union Rules the World (OUP 2020).

## 2. A Typology of Data Flows Grounded in Prevailing Practice
### 2-1. Purpose and Scope of This Chapter

This chapter develops a typology of how data flows are structured in practice. Although this paper examines how to determine the law that should govern private-law disputes arising from cross-border data transfers, the assumed fact pattern is consequential in two respects.

First, it may affect the dispute's legal characterization, in particular whether liability is framed in contract (breach of contract) or outside of contract (i.e. tort). As discussed later, this paper argues for the need for private ordering and seeks a private international law framework capable of addressing these two bases of liability in a unified manner. The typological analysis undertaken in this chapter provides an essential foundation for that argument. In particular, given the prevailing practice of relying on privacy policies, a key consideration is the potential concurrence of a claim grounded in breach of contract and a claim grounded in tort.

Demonstrating this requires identifying the relevant actors and clarifying the legal relationships among them, especially whether a direct contractual relationship exists.

Second, the assumed fact pattern is relevant to the technical methods employed for the storage, processing, and use of data. In particular, this chapter discusses secret sharing techniques for distributing data across multiple locations for storage. This serves to demonstrate that the traditional approach of private international law, which determines the applicable law by reference to a physical location, does not function adequately in the context of cross-border data transfers, thereby providing the groundwork for the subsequent argument in favor of the necessity of private ordering.

Accordingly, this chapter serves a dual function: it concretizes the problem domain through a practice-oriented typology and supplies the considerations necessary for the legal-theoretical analysis developed in the chapters that follow.

### 2-2. Overview of the typology of data flows: clarifying the actors and their legal relationships

This part then provides an overview of the typology of data flows, illustrated in Figure 1, to clarify the actors and the legal relationships among them. As this paper examines the choice-of-law frameworks under EU law and Japanese law, the following typology is organized, for the sake of analytical convenience, with reference to data flows between an EU Member State (hereinafter "EU MS") or Japan on the one hand, and a third country on the other.

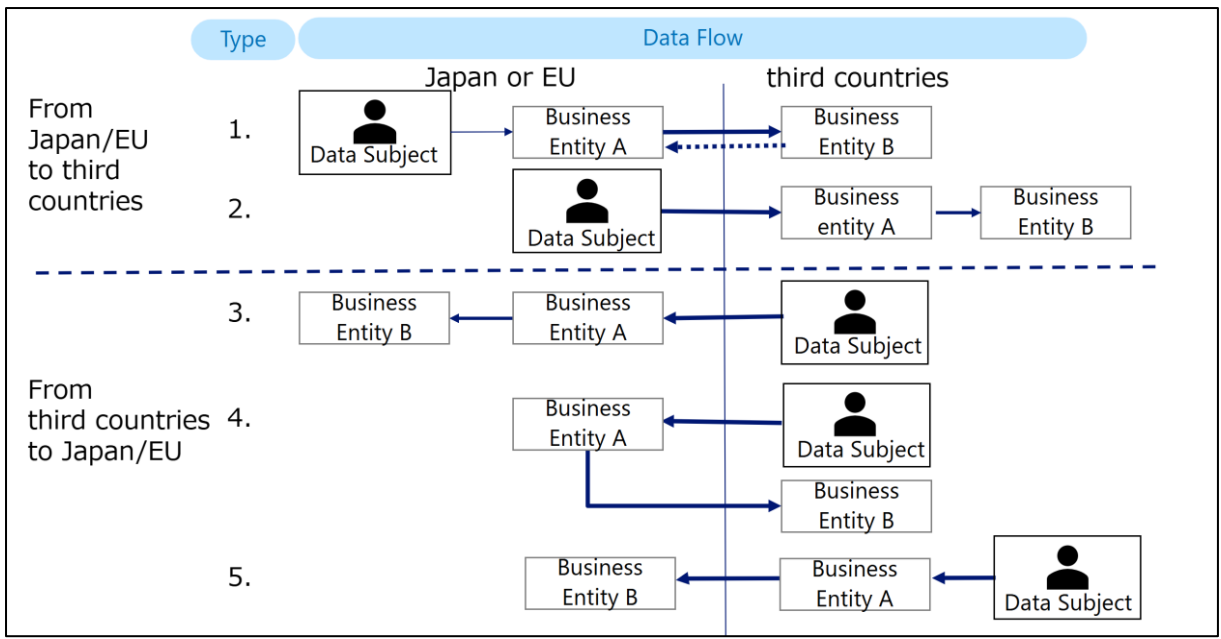

**Figure 1**: Types of Data Flows

It is appropriate to begin by identifying the relevant actors, to build a foundation for understanding their relationships.

First, the "Data Subject" is the source of the data that business operators obtain, disclose, and use, and is assumed, in principle, to be a natural person such as a consumer. Admittedly, the term "data subject" is found in data protection law such as Article 4(1) of the GDPR and has a broad scope of application. In this paper, however, the Data Subject is assumed, in principle, to be a user (essentially a consumer) who utilizes the services of Business Entity A, as explained below.

Second, "Business Entity A" is an entity that provides some service to the Data Subject directly. This includes, for example, shopping websites and AI chat services—so-called SaaS, as explained below—i.e., entities with which the Data Subject, as a consumer, can recognize that they have an active and direct point of contact, and with which the Data Subject typically has a direct contractual relationship.

Third, "Business Entity B," by contrast, is an entity that provides the technical means necessary for Business Entity A to deliver such services. Examples include providers of cloud infrastructure (such as Amazon AWS)—so-called IaaS or PaaS, as explained below—used to deliver applications in the cloud, as well as providers of data centers that store data related to the services. The Data Subject, as a consumer, is not necessarily aware that such services or facilities are being used; accordingly, there is typically no direct contractual relationship between the Data Subject and Business Entity B.

With these actors in mind, the table below sets out a schematic description of each type illustrated in Figure 1, together with a practical account of how it operates in real-world settings.

Table 1. Types of Data Flows Involving Cross-Border Transfers and their Explanations

| Type | Description | Use Case |
|---|---|---|
| 1 | Business Entity A (in Japan or the EU) collects personal data from Data Subjects and transfers it to Business Entity B (in a third country) for analysis and processing. | ・Using overseas cloud services (e.g. AWS, Google Cloud) to process data collected in Japan or the EU (and, in practice, possibly returning the processed data to Japan or the EU)<br>・Japanese users making reservations at foreign hotels, etc. through Japanese travel agencies and registering personal information |
| 2 | Data Subject (in Japan or the EU) provides personal data directly to Business Entity A (in a third country), which further transfers it to Business Entity B (also in a third country). | ・Japanese or EU users making reservations and registering personal information on websites for hotels located overseas (an example of extraterritorial application under Article 171 of Japanese Act on the Protection of Personal Information or Article 3 of GDPR) |
| 3 | Data Subject (in a third country) provides personal data to Business Entity A (in Japan or the EU), which transfers it to Business Entity B (also in Japan or the EU). | ・Overseas users make reservations and register personal information for Japanese events and exhibitions |

| 4 | Data Subject (in a third country) provides personal data to Business Entity A (in Japan or the EU), which then transfers it to Business Entity B (in a third country). | ・Overseas users register personal information on servers in Japan or the EU, which is subsequently utilized by overseas businesses |
|---|---|---|
| 5 | Data Subject (in a third country) provides personal data to Business Entity A (also in a third country), which further transfers it to Business Entity B (in Japan or the EU). | ・Acquiring personal information of overseas subsidiary employees at the Japanese or EU parent company<br>・Transferring overseas log information to Japan or EU |

In light of the types of data flows outlined above, when considering claims arising from the leakage of personal data due to cyberattacks or similar incidents, it is appropriate to take into account the following relationships for each type.

To this end, this section first identifies features common to all types before turning to an examination of each type individually.

As a preliminary point, it should be noted that the Data Subject and Business Entity A are linked by a direct contractual relationship. For example, a consumer (=Data Subject) may enter into a contract to use the services provided by a SaaS provider (=Business Entity A). In addition, a direct contractual relationship also exists between Business Entity A and Business Entity B. For example, a SaaS provider (=Business Entity A), in offering its applications to users, may utilize cloud infrastructure for deploying those applications and enter into a contract with an IaaS provider (=Business Entity B) for that purpose. It is also conceivable that Business Entity A may contract with a data center operator (=Business Entity B) for the storage of data.

Accordingly, where the Data Subject brings a civil claim for compensation against Business Entity A, or where Business Entity A brings such a claim against Business Entity B—for instance, where Business Entity B is a data center operator and data is leaked due to inadequate security measures, causing a loss of user trust and resulting damage to Business Entity A—, the claim may be grounded in breach of contract. Moreover, since contractual liability and statutory tortious liability constitute separate bases of civil liability, the existence of a contractual relationship does not necessarily preclude liability in tort. It is therefore also possible to bring a civil claim for compensation on the basis of tort.

By contrast, there is no direct contractual relationship between the consumer (=Data Subject) and Business Entity B. Even if Business Entity A includes a provision in its privacy policy, at the time of contracting with the Data Subject, stating that it utilizes infrastructure provided by Business Entity B, such a policy is prepared solely by Business Entity A. Since Business Entity B is not a party to that contract, the privacy policy does not give rise to a contractual relationship between the Data Subject and Business Entity B.

Therefore, in the above example of data leakage, where the Data Subject seeks to bring a civil claim for compensation against Business Entity B for having tolerated inadequate security measures, the claim cannot be grounded in contractual liability, and must instead be based solely on tortious liability.

With these common features in mind, the discussion now turns to an examination of each type. At this point, although one might envisage examining each type individually and, such an approach is not appropriate for the following reasons. The types set out above represent a typological classification of the diverse data flows that arise in actual economic activity, organized according to the geographical location of the Data Subject, Business Entity A, and Business Entity B, respectively. Accordingly, taking as a starting point the common legal situation regarding the existence or absence of contractual relationships described above, the analysis should focus on which claims between which actors involve foreign elements. Viewed in this light, rather than examining each type one by one, it is more appropriate to classify claims into those that involve a foreign element and those that do not. From this perspective, the claims in each type can be classified as shown in Table 2 below. It should be noted that the contents of this Table are referred to in the subsequent discussion in the present paper and accordingly, a reference number has been included in the leftmost column for convenience of citation.

Table 2: Classification of Claims of Each Type by Presence or Absence of a Foreign Element

A. Claims **involving** foreign elements

| No. | type | Claimant→Respondent | Basis of Liability |
|---|---|---|---|
| ① | 1, 5 | Data Subject → Business Entity B | Non-contractual |
| ② | 1, 5 | Business Entity A → Business Entity B | Contractual and non-contractual |
| ③ | 2, 3 | Data Subject → Business Entity A | Contractual and non-contractual |
| ④ | 2, 3 | Data Subject → Business Entity B | Non-contractual |
| ⑤ | 4 | Data Subject → Business Entity A | Contractual and non-contractual |
| ⑥ | 4 | Business Entity A → Business Entity B | Contractual and non-contractual |

B. Claims **not** involving foreign elements

| No. | type | Claimant→Respondent | Basis of Liability | Governing Law |
|---|---|---|---|---|
| ⑦ | 1, | Data Subject → Business Entity A | Contractual and non-contractual | Japanese or EU Law |
| ⑧ | 5 | Data Subject → Business Entity A | Contractual and non-contractual | Law of the Third Country |
| ⑨ | 2 | Business Entity A → Business Entity B | Contractual and non-contractual | Law of the Third Country |
| ⑩ | 3 | Business Entity A → Business Entity B | Contractual and non-contractual | Japanese or EU Law |
| ⑪ | 4 | Data Subject → Business Entity B | Non-contractual | Law of the Third Country |

     As demonstrated in this section, data flows in the real world can be systematically organized by first classifying the relevant actors and then examining the presence or absence of contractual relationships among them and of foreign elements in the resulting claims. In light of the nature of the present paper — which examines international civil liability in the context of cross-border data transfers — scenarios ① through ⑥ constitute the subject of analysis.

## 2-3. Cloud Computing and Secret Sharing Techniques: The Diminishing Significance of Physical Location

### 2-3-1. Cloud Computing as Contrasted with On-Premises Infrastructure

Until the widespread adoption of cloud computing in the late 2000s, most service providers installed their own servers and storage at their own facilities in their home countries to provide services (Figure 2 (1)). In addition to the operation of applications and databases necessary for service provision, various data generated and stored through service user registration and usage were also stored and managed on these servers and storage. Furthermore, the companies themselves were primarily responsible for monitoring server operations, responding to failures, and expanding server and storage capacity in response to increasing users and service expansions.

Following the widespread adoption of cloud computing, many business entities began using it, citing benefits such as reduced monitoring and operational costs and improved availability in the event of failures (Figure 2 (2)). Cloud computing offers advantages such as improved fault tolerance and enhanced availability through distributed data management, by distributing servers and storage across multiple locations[13]. Figure 2 conceptually illustrates the differences from On-Premises.

When classifying cloud services by function, there are three main types[14]: SaaS (Software as a Service), which provides the consumer with the capability to use the provider's applications running on a cloud infrastructure; PaaS (Platform as a Service), which provides the capability to deploy onto the cloud infrastructure consumer-created or acquired applications created using programming languages, libraries, services, and tools supported by the provider; and IaaS (Infrastructure as a Service), which provides the capability to provision processing, storage, networks, and other fundamental computing resources where the consumer is able to deploy and run arbitrary software, including operating systems and applications. SaaS includes services directly used by end-users, such as shopping sites and social networking services, while IaaS and PaaS are primarily provided to service providers.

As noted above, SaaS corresponds to Business Entity A, and PaaS or IaaS corresponds to Business Entity B, both as used in the typological framework in the preceding chapter.

---

[13] Cloud service providers generally implement control measures to meet the information security requirements of ISO 27001 and ISO 27017.

[14] Mell and Grance (n 4).

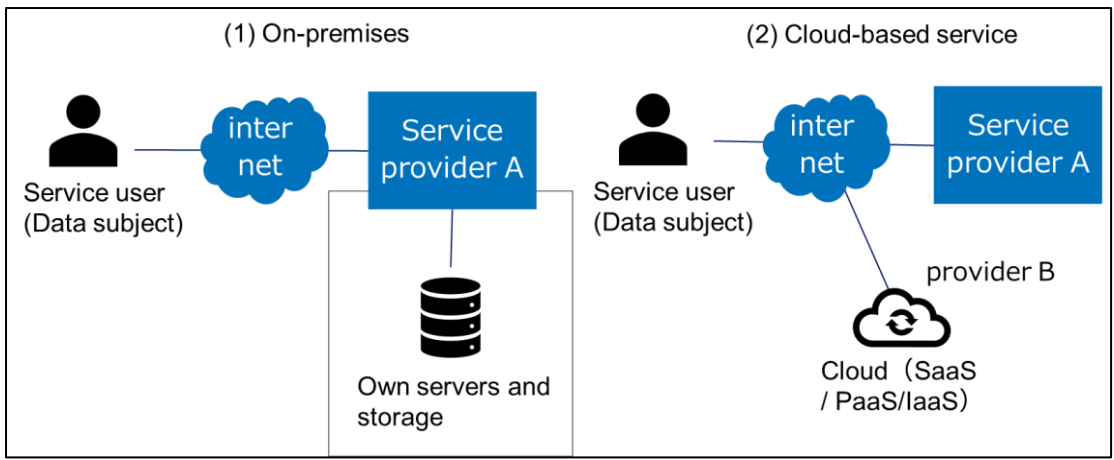

Figure 2. On-Premises vs. Cloud-Based Services

### 2-3-2. Features of Cloud Computing from a Data Distribution Perspective

From a data distribution and storage perspective, when using on-premises servers, the server's location is roughly clear down to the street and address level, making it clear where and in which country the data is stored and processed. On the other hand, with cloud services, as a security threat countermeasure, the exact location is often not disclosed, with only a general region (Tokyo, Osaka, San Francisco, etc.) or, at most, a broader geographical area (Asia, Japan, South Korea, Europe, etc.) being indicated[15].

For the users, not only is the location of the data unknown, but there is also the possibility that data may be transferred from one's own country to other countries without the user's knowledge. Cloud providers may install servers across multiple countries, and for fault tolerance and backup purposes, identical copies of data may be replicated and stored across multiple servers in different locations[16]. In this paper, this type of distributed system called as the "replication type system".

---

[15] Cloud service providers such as AWS do not disclose the exact location of their servers, instead indicating their location at the regional level.
See eg Amazon Web Services, 'Regions, Availability Zones, and Local Zones' <https://docs.aws.amazon.com/AmazonRDS/latest/UserGuide/Concepts.RegionsAndAvailabilityZones.html> accessed 30 March 2026.

[16] As a foundational study on improving fault tolerance of systems through replication, see Leslie Lamport, Robert Shostak and Marshall Pease, 'The Byzantine Generals Problem' (1982) 4(3) ACM Transactions on Programming Languages and Systems 382. For replication techniques that improve fault tolerance through distributing data across multiple nodes in the context of cloud computing, see  Giuseppe DeCandia and others, 'Dynamo: Amazon's Highly Available Key-Value Store' (2007) 41(6) ACM SIGOPS Operating Systems Review 205.

Furthermore, over the past 20 years, technologies such as encryption and secret sharing[17] have been developed to protect important or sensitive data, dividing it into multiple fragments and managing them on separate servers, leading to progress in distributed data management. This paper shall refer to this type of distributed system as the "secret sharing type system".

In a replication type system, compromising a single node is sufficient to obtain the complete data, as each node holds an identical copy. By contrast, in a secret sharing type system, the data is divided into *n* shares such that at least *k* shares (where $k \leq n$) are required to reconstruct the original data, and any fewer than k shares reveal no information about the original data[18]. Here, *k* is called the threshold. A higher *k* increases confidentiality against attacks, as an attacker must compromise more nodes to reconstruct the data. However, a higher *k* also reduces availability, since more nodes must be operational to reconstruct the data. This trade-off between confidentiality and availability means that an appropriate threshold is determined for each system based on its specific requirements. Regardless of the threshold chosen (provided $k \geq 2$), the key distinction from a replication type system is that compromising a single node alone is insufficient. These characteristics are illustrated in Figure 3 below, which incorporates a cross-border element by locating each node in different countries.

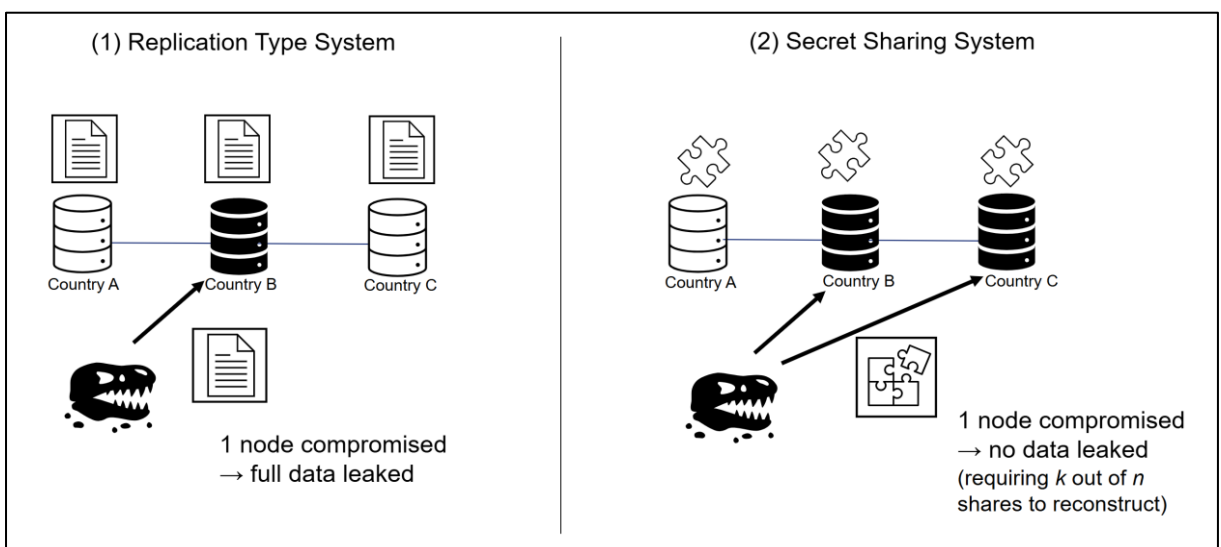

Figure 3. Replication Type System vs. Secret Sharing Type System: Contrasting Features

---

[17] Adi Shamir, 'How to Share a Secret' (1979) 22(11) Communications of the ACM 612.
[18] See Shamir (n 17).

As discussed above, while recent cloud adoption has improved operability and availability for both service users (i.e., data subjects) and service providers, it also presents the problem of difficulty in understanding and identifying the location of data. Hence, if data is leaked due to a cyberattack or unauthorized data theft at a given cloud location, it becomes difficult to determine the location of the leak. This difficulty becomes particularly acute in secret sharing type systems, where data is fragmented and distributed across multiple nodes.

Moreover, the significance of secret sharing extends beyond mere data storage and protection. In recent years, it has in particular become a fundamental technology for secure computation, which enables processing or computation on data without revealing it to any single party. Secure computation has already found practical applications in fields handling sensitive data, such as health care[19], and it is expected to play an increasingly important role in the context of training AI models[20]. Hence, taking this technology into account in building a legal framework concerning cross-border data transfers is essential for both present and future AI-driven society.

## 3. Choice-of-Law Issues in the context of Cross-Border Data Transfer: Cloud Computing as a Challenge to the Principles of *lex loci damni* and *lex loci delicti commissi*

### 3-1. Introductory Remarks

The classification of data flows from a technical perspective, as shown in the previous section, suggests that there will be an increase in cases that are difficult to address using traditional choice-of-law principles on non-contractual liability. Specifically concerning the use of Secret Sharing Type System, in which data fragments are stored across borders as described in Figure 3, neither the principle of "lex loci damni" nor that of "lex loci delicti commissi" seems to be appropriate in terms of the parties' foreseeability.

---

[19] Koji Chida and others, 'Implementation and Evaluation of an Efficient Secure Computation System Using "R" for Healthcare Statistics' (2014) 21 Journal of the American Medical Informatics Association 326.

[20] Payman Mohassel and Yupeng Zhang, 'SecureML: A System for Scalable Privacy-Preserving Machine Learning' (2017 IEEE Symposium on Security and Privacy, San Jose, May 2017) 19; Nuttapong Attrapadung and others, 'Adam in Private: Secure and Fast Training of Deep Neural Networks with Adaptive Moment Estimation' (2022) 2022(4) Proceedings on Privacy Enhancing Technologies 746.

As a preliminary to examining this issue, the following provides an overview of the choice-of-law rules under EU law and Japanese law. Here, two observations should be noted before proceeding.

First, regarding the treatment of the GDPR in the illustration of EU law. In fact, GDPR contains no rule pertaining to the applicable substantive data-protection law. Despite the GDPR's character as a directly applicable Regulation, determining the applicable law is essential, because GDPR incorporates numerous opening clauses (e.g., Articles 6(2) and (3)[21], 8(1)[22], 9(4)[23], 23,[24] 85[25], 88[26] GDPR). Opening clauses afford Member States regulatory discretion, thereby producing divergent national legal regimes across the Union. Proceeding from the assumption that Article 82 constitutes a tortious claim[27] that despite the wording in

---

[21] Article 6(2) and (3) contain opening clauses that enable Member States and the Union to maintain or introduce sector-specific data protection rules. The regulatory discretion is narrow in scope, as it extends only to processing operations based on Article 6(1)(c) and (e).

[22] Article 8(1) contains an opening clause allowing Member States to determine, within a range between the ages of 13 and 16, the point at which children may independently consent to the processing of personal data in relation to information-society services. E.g. Austria has made use of this discretion, see § 4(4) DSG.

[23] Article 9(4) grants Member States the discretion to maintain or introduce conditions pertaining to the processing of genetic data, biometric data or health data.

[24] The provision grants Member States a broad discretion to restrict any of the data-subject rights. While neither the Regulation nor Recital 73 delineates concrete conditions, both anchor their permissibility in general constitutional considerations that any limitation "*respects the essence of the fundamental rights and freedoms and is a necessary and proportionate measure in a democratic society to safeguard,*" particularly to secure essential objectives of general public interest within the Member State concerned.

[25] Article 85(1) grants Member States the competence to fine-tune, through statutory provisions, the normative relationship between data protection and freedom of expression.

[26] The opening clause contained in Article 88, affords Member States the competence, within the confines of the provision, to develop national data protection regimes for the employment context.

[27] Note that the right to an effective judicial remedy is not limited to compensatory claims under Article 82 GDPR. However it is evident from case law of CJEU that this article provides for a statutory right of data subject specific to the GDPR (see Case C-300/21 *UI v Österreichische Post AG* EU:C:2023:370; Václav Janeček and Cristiana Teixeira Santos, 'The Autonomous Concept of "Damage" according to the GDPR and Its Unfortunate Implications: Österreichische Post' (2024) 61 CMLR 531).

Article 1(2)g of the Regulation is not exempt from Rome II,[28] the applicable law would, generally be the law of the place where the damage occurs[29]).

Second, regarding the necessity of distinction between contractual liability and tortious liability. Processing operations concerning the personal data of natural persons frequently occur in transnational constellations, thereby prompting the question of which national damages regime governs the claims of affected individuals. This necessitates a differentiation between contractual and tortious bases of liability. This point is illustrated in Table 2 (Section 2 of this paper), which examines the content and legal nature of claims between each actor in the context of cross-border data transfers, as well as the presence or absence of foreign elements.

### 3-2. Overview of EU law and Japanese law concerning choice of applicable law
### 3-2-1. Contractual Liability

Within the EU, the Rome I Regulation[30] establishes the parties' autonomy to select the law applicable to contractual obligations. The parties may freely determine the national law governing their contract, a choice that also encompasses ancillary issues, including claims for damages.[31] Article 6(2) Rome I, restricts the parties' autonomy to choose the applicable law in consumer contracts to the extent that the consumer continues to enjoy the protection of the mandatory provisions of the law that would have governed the contract in the absence of such a clause. According to Article 6(1) Rome I, a contract concluded between a consumer and a professional (B2C) is governed by the law of the consumer's habitual residence, provided that the professional pursues his commercial or professional activities in that country or, inter alia, directs such activities to it, and the contract falls within the scope of those activities. Article 6(4) Rome I precludes certain types of contracts[32], such as service contracts performed exclusively outside the consumer's State, from this protective regime. These contracts are governed by the law that would be applicable to non-consumer contracts.
In the absence of a (valid) choice of law-clause in the contract, the applicable law for contractual relationships within the EU must be determined pursuant to Article 4 Rome I Regulation. With respect to claims for damages, this leads to the national law governing the principal contract, provided that the processing of personal data does not constitute the primary object of the contractual relationship. In contracts of sale[33] or service[34] contracts, the applicable law is that of the seller's or service provider's habitual residence. In contracts concerning rights in rem or leases and tenancies of immovable property exceeding six months, the lex rei sitae applies.[35] Where the contract type giving rise to the damages claim is not enumerated in Article 4(1) Rome I, the characteristic performance is decisive, and the applicable law is that of the habitual residence of the party effecting that characteristic performance.[36] Finally, the fallback clause in Article 4(4) Rome I provides that, where the applicable law cannot be determined under paragraphs 1 or 2, the law of the country most closely connected to the contract shall apply.

---

[28] See discussion under 4.6.2.2.

[29] Leupold and Schrems argue that neither Rome I nor Rome II apply. Instead, they purport that in the absence of a special rule, the applicable national conflict-of-laws rules of the lex fori apply. *Leupold/Schrems* in *Knyrim*, DatKomm Art 79 DSGVO (1.3.2021, rdb.at) para 57. Schweiger supports this perspective. *Schweiger* in *Knyrim*, DatKomm Art 82 DSGVO (1.12.2021, rdb.at), para 84. Controversial, *Becker* in *Plath*, Kommentar zu DSGVO, BDSG und TTDSG (4th Edition, 2023), Article 82, para 17.

[30] Regulation (EC) 593/2008 of the European Parliament and of the Council of 17 June 2008 on the law applicable to contractual obligations (Rome I), OJ 2008 L 177/6.

[31] See Article 3(1) Rome I.

[32] See Article 6(4) Rome I.

Under Japanese private international law, issues of contractual liability are governed by the law applicable to contractual matters, regardless of whether the parties sue for breach of contract or for tort (Article 7 or Article 20 of AGRALJ). As for the law applicable to contractual matters, it is, in principle, determined through the party autonomy principle (Article 7 of AGRALJ). If the parties do not agree upon the choice-of-law for their contract, the law of the place most closely related to the contract shall apply (Article 8 of AGRALJ).

### 3-2-2. Tortious Liability

Proceeding on the assumption that Article 82(1) constitutes a tort-like claim for damages, the initial question would be whether the Rome II Regulation is applicable. Non-contractual obligations arising from infringements of privacy or rights relating to personality fall outside the scope of the Rome II.[37] Schweiger understands that this precludes the application of Rome II in the context of GDPR infringements.[38] Becker, contemplating the intention of Article 1(2)(g) Rome II, maintains that the exception should be interpreted narrowly. He argues that Article 1(2)(g) Rome II was borne by the consideration that law provisions related to media and tort-based injunctions have not yet been harmonized at Union level, with the result that international jurisdiction in such matters should, for the time being, remain governed by national conflict-of-laws rules. Conversely, the exemption should not encompass claims under Article 82 GDPR. The law applicable would then be determined primarily pursuant to Article 4(1) Rome II, that is, by the law of the Member State in which the damage occurs, irrespective of the State in which the event giving rise to the damage or any indirect consequences took place. Typically, this will be the law of the Member State in which the data subject has his or her habitual residence.[39] On this issue, determining the law applicable to non-contractual liability in this context will be further discussed in the next section.

---

[33] Article 4(1)(a) Rome I.

[34] Article 4(1)(b) Rome I.

[35] Article 4(1)(c) Rome I.

[36] Article 4(2) Rome I.

[37] See Article I(2)(g) Rome II.

[38] *Schweiger* in *Knyrim*, DatKomm Art 82 DSGVO (1.12.2021, rdb.at), para 84.

Under Japanese private international law, AGRALJ Article 17 governs torts in general (*lex generalis*), while two special provisions (*lex specialis*) govern product liability (Article 18) and defamation (Article 19). Article 17, in principle, relies on the connecting factor of "the place where the result of the wrongful act occurred." However, in case it is normally impossible for reasonable persons to foresee the occurrence of the result at the relevant place, "the law of the place where the wrongful act was committed" will govern (the second sentence of AGRALJ Article 17). In case of defamation, Article 19 adopts the victim's habitual residence as its connecting factor. Considering difficulties to properly apply these provisions for liability issues among parties of international data management contracts, the next section will elaborate the potential role of escape clause in this context.

### 3-3. Difficulty of determining the applicable law based on the geographical territorial element in the context of cloud computing

As for contractual liability, the applicable law is, in principle, determined by party autonomy under both EU law and Japanese law (Article 6 of the Rome I Regulation; Article 7 of AGRALJ). Accordingly, the technical characteristics of cloud computing — namely, the difficulty of identifying a physical location — do not, in themselves, give rise to an immediate problem; the parties need only agree upon the law of any given jurisdiction as the applicable law, without being bound by the identification of a physical location. Hence, with respect to claims ②, ③, ⑤, and ⑥, insofar as they are grounded in contractual liability, the recourse concerning the determination of the applicable law can, in principle, be found through party autonomy.

In contrast, as regards the law applicable to tortious liability as a non-contractual obligation both EU law and Japanese law establish the principle that the law of the place where the damage occurs shall govern (Article 17 of AGRALJ; Article 4(1) of the Rome II Regulation).

---

[39] *Becker* in *Plath*, Kommentar zu DSGVO, BDSG und TTDSG (4th Edition, 2023), Article 82, para 17.

As noted above, under EU law, the place where the damage occurs has been interpreted as the place where the data subject has his or her habitual residence; however, this interpretation is not explicitly set forth in the text of the provision, and alternative views are conceivable. For instance, the place where data was leaked as a result of a cyber-attack could also be regarded as the place where the damage occurred.

Under Japanese law, with the exception of defamation cases (Article 19 of AGRALJ), there is no provision stipulating that the place where the damage occurs corresponds to the place where the data subject has habitual residence, nor is there any prevailing scholarly opinion to that effect; accordingly, the determination must be made on a case-by-case basis. One possible interpretation is that the location of the data leak may serve as the connecting factor to determine applicable law, as well as under EU law.

Furthermore, unlike EU law, Japanese law contains an explicit provision whereby, in case it is normally impossible for reasonable persons to foresee the occurrence of the result at the relevant place, the law of the place where the wrongful act was committed shall govern (the second sentence of AGRALJ Article 17). Where a data leak results from a cyber-attack, it would appear to be a reasonably straightforward and tenable interpretation that the place where such an attack was carried out falls within this provision.

However, as noted above, even in the case of a replication-type system, the exact location remains undisclosed from a security standpoint. Even if the disclosure of such information is to be demanded, this could expose the data of numerous users relying on the cloud service to risk, with the consequence that such a request may be lawfully refused pursuant to civil procedure law. In the case of a secret-sharing-type system, furthermore, the identification of the leak location is technically difficult in the first place, given that the information is fragmented and distributed across multiple locations. Even if the node that served as the entry point for the cyber-attack can be identified — for instance, through log analysis revealing anomalous activity at a particular node — the remaining $n$-1 nodes, each holding a fragment of the data distributed under a secret sharing scheme, would still serve as candidates, which may be spread across multiple jurisdictions, rendering it impossible to determine a single location.

As demonstrated above, the traditional private international law approach of determining the applicable law on the basis of a physical location presents considerable difficulties in the choice of law governing cross-border data transfers, under both Japanese law and EU law, particularly with respect to non-contractual obligations. More specifically, as illustrated in Table 2 in chapter 2 of this paper, these difficulties are especially acute in relation to claims ① and ④; moreover, the same problems arise in respect of claims ②, ③, ⑤, and ⑥, insofar as they are pursued on the basis of non-contractual obligations. It is therefore necessary, in these cases, to determine the applicable law from an alternative perspective, departing from the traditional reliance on the identification of a physical location.

## 4. Potential of Party Autonomy as a Means of Determining the Governing Law of Non-Contractual Obligations

### 4-1 Introductory Remarks

The question then arises as to what method would be appropriate for determining the governing law applicable to claims based on such non-contractual obligations. In order to undertake this examination, it is useful to recall Table 2 and to identify the specific types of civil disputes that are likely to arise in practice.

At the outset, a Data Subject who has suffered damage as a result of a data leak—which may, in some cases, give rise to serious harm such as defamation or the disclosure of trade secrets — would bring a claim for damages against Business Entity A (corresponding to a SaaS provider), which the Data Subject can identify as its contractual counterpart because of using their service directly as illustrated above. In such a scenario, Business Entity A would normally include a clause stipulating the exclusion of liability in its privacy policy or general terms and conditions (*allgemeine Geschäftsbedingungen*) and then, it may lead the Data Subject to ground the claim in tortious liability — a form of non-contractual liability — rather than contractual liability. If this claim is upheld, and the data leak is found to have originated from a vulnerability in the infrastructure provided by Business Entity B (corresponding to a PaaS or IaaS provider), Business Entity A may in turn bring a claim against Business Entity B for damages, encompassing both the compensation paid to the Data Subject and the resulting loss of reputation.

As is evident from the discussion above and as illustrated by the "Basis of Liability" column for claims ③ and ⑤, as well as ② and ⑥ in Table 2, the typical civil disputes that arise most frequently in practice tend to involve a concurrence of contractual and tortious liability. This stands in contrast to the categories represented by ① and ④, where no direct contractual relationship exists and tortious liability alone constitutes the basis of the claim. Accordingly, it is appropriate to examine choice-of-law measures capable of overcoming the limitations associated with the identification of a physical location in the context of cloud computing, having due regard to the distinctive features of such cases.

### 4-2. Party Autonomy for Accommodating Private Ordering

In the field of non-contractual claims, the role of party autonomy has traditionally been discussed prudently, recognizing the asymmetrical power dynamics in negotiations between contracting parties.[40] Even though it is generally accepted that parties may agree on the applicable law of tort after the occurrence of the tortious event,[41] both EU and Japanese law take a cautious stance on whether prior agreement to choose the governing law of tort is permissible. Under the Rome II Regulation, a prior agreement about the law applicable to tortious claims could be concluded only "where all the parties are pursuing a commercial activity, also by an agreement freely negotiated before the event giving rise to the damage occurred."[42] A restrictive attitude toward prior agreement on the governing law of tort based on similar considerations is more clearly evident in Japanese law. The AGRALJ provides no provision regarding prior consent to the law applicable to tortious claims, and it is generally understood that such prior agreement could not be validly given, even between parties engaged in commercial activities.

However, it would be inaccurate to interpret this legislative trend as negating the role of prior agreement for non-contractual claims in its entirety, given that both the Rome II Regulation and the AGRALJ in Japan include escape clauses. Rather, in recent years, in the context of international tort litigation between online platform operators and their service users, there has been a noteworthy argument that the existence and effective use of the escape clause can lead to respect for private ordering between the parties.[43] From the perspective of this argument, the following paragraphs will examine how choice-of-law issues on non-contractual claims classified in Table 2 should be resolved.

In cases corresponding to ③ or ⑤ in Table 2, in which the Data Subject makes a lawsuit against the Business Entity A,[44] it is admittedly usual that the platform contract between them contains a choice-of-law clause.[45] Also, if there is a choice-of-law clause in their contract, allowing the same chosen law to govern both contractual and non-contractual claims would increase certainty and foreseeability for the parties, including both the Data Subject and Business Entity A.[46]

Accordingly, if a court in an EU Member State takes jurisdiction over the cases ③ or ⑤ in Table 2, in the most part of non-contractual claims, the court should be able to take the pre-existing contractual relationship between the Data Subject and Business Entity A, resulting in the application of the same law to their contractual and non-contractual issues pursuant to Article 4 (3) of the Rome II Regulation. Although there is a considerable argument that the law designated through Article 4 (3) should be different from the law of the place where damages occurs,[47] considering the structural obstacle to point a single jurisdiction as the place of damage in tort cases of international data management, it would be desirable to allow the court to make use of the escape clause without such requirement.

---

[40] See, for example, Mo Zhang, 'Party Autonomy in Non-Contractual Obligations: Rome II and Its Impacts on Choice of Law' (2009) 39(3) Seton Hall Law Review 861-917.

[41] See Article 14 (1) (a) of the Rome II Regulation. See also Article 21 of the AGRALJ.

[42] See Article 14 (1) (b) of the Rome II Regulation.

[43] Tobias Lutzi, Private International Law Online: Internet Regulation and Civil Liability in the EU (OUP 2020) 169–183.

[44] Admittedly, the classification presented in Table 2 is not fully effective for discussing international jurisdiction, which shall be discussed in a separate paper.

[45] See Lutzi (n 43)170.

[46] See Lutzi (n 43)171.

On the other hand, as to the issue of defamation, which is not covered by the Rome II Regulation, the governing law should be determined through the choice-of-law rules in the forum State. While how to determine the governing law of defamation is a matter left to each member state, it would be desirable for the choice-of-law process to take into account the practicalities of private ordering in the field of personal data protection.

If a lawsuit of the type ③ or ⑤ in Table 2 is brought in a Japanese court, the court may take into account their pre-existing relationships, even for determining the law applicable to non-contractual issues. For determining the law applicable to the damage caused by tortious acts, Article 17 of the AGRALJ provides a general rule, under which "the place where the result of the wrongful act occurred" is a connecting factor in the first sense (the first sentence of that Article), but the law of "the place where the wrongful act was committed" overrides when the "occurrence of the result at the relevant place was ordinarily unforeseeable" (the second sentence of that Article). Either of the connecting factors established by the first or second sentence of AGRALJ Article 17 would cause difficulties in its application, especially in cases where the structure of secret-sharing technology is established internationally as above discussed. However, even in such cases, the Japanese court may rely on AGRALJ Article 20 as an escape clause, which allows the court to apply "the law of the place with which the tort is obviously more closely connected" than the place indicated in AGRALJ Articles 17-19. AGRALJ Article 20 explicitly states that, under this provision, the court may consider pre-existing contractual relationships between the parties.

If a claim on defamation matters in a Japanese court, the choice-of-law rule for defamation is set out in Article 19 of the AGRALJ, which is a *lex specialis* in relation to Article 17 of the AGRALJ[48]. Speaking of the material scope of Article 19 of the AGRALJ, this provision covers the violation of a person's reputation or business credibility, which is narrower than the personality rights in general[49]. Accordingly, wrongful acts against other personality rights, except for a person's reputation or business credibility, are covered by Article 17 of the AGRALJ[50]. In the case discussed here, the alleged tort should be regulated by the law determined through Article 19 of the AGRALJ in principle.

---

[47] See Richard Fentiman, 'The Significance of Close Connection' in John Ahern and William Binchy (eds), The Rome II Regulation on the Law Applicable to Non-Contractual Obligations: A New International Litigation Regime (Martinus Nijhoff 2009) 85, 89.

[48] In the legislative process of the AGRALJ, several more special connecting factors on torts were envisaged such as "the country of protection for the infringement of intellectual property rights"

Then, since the connecting factor stipulated by Article 19 is the "victim's habitual residence (in cases where the victim is a juridical person or any other association or foundation, the law of its principal place of business)," in the cases like ③ or ⑤ in Table 2, applying the substantive law of the Data Subject's habitual residence would be a candidate solution. However, this solution would lead to the result that the governing law on defamation arising from the mismanagement of personal data can vary depending on the habitual residence of the potential victims, even though the alleged harmful act has occurred in relation to the same data management contract. Accordingly, concern may arise about the foreseeability of choice-of-law issues, especially as the number of Data Subjects contracting with Business Entity A increases.

Then, would it also be acceptable to consider Article 20 of the AGRALJ in this context? As mentioned above, AGRALJ Article 20 is a provision to allow the court to apply "the law of the place with which the tort is obviously more closely connected" than the place indicated in AGRALJ Articles 17-19. This provision exemplifies two situations in which this escape clause should be invoked. One is the situation in which "the parties had their habitual residence in the places governed by the same law at the time of the occurrence of the tort," and the other is the situation where "the tort was committed in breach of the obligation under a contract between the parties." These are illustrative, not restrictive, enumerations. In situations where a contract regarding data management exists, if defamation arises from data management matters, it would be difficult to see it as falling directly under the latter exemplified pattern. However, from the perspective of foreseeability, reliance on the escape clause in such a situation should be legitimate.[51]

As for cases corresponding to ② or ⑥ in Table 2, where the Business Entity A

---

or "the marketplace for acts violating antitrust law," but they were not adopted in AGRALJ. See Kazuaki Nishioka and Yuko Nishitani, Japanese Private International Law (Hart Publishing 2021) 124–125.

[49] Jun Yokoyama, Private International Law in Japan (Wolters Kluwer, 2019) 78.

[50] Ibid.

[51] See Tokyo District Court Judgement, 15 October 2024, as an example of a case where it would have been desirable to rely on AGRALJ Article 20 in order to determine the applicable law for defamation. However, in this judgment, the court failed to clearly refer to the escape clause.

corresponding to SaaS provider) makes a lawsuit against the Business Entity B (Corresponding to IaaS or PaaS provider), basically, the above discussion in this section is relevant to the significance of making use of the escape clauses for the court to align the governing law of non-contractual issues with that of contractual issues, especially when there is a choice-of-law agreement between A and B. However, there may be cases where, if Business Entity A is required to compensate a Data Subject, Business Entity A seeks partial reimbursement from Business Entity B. In such a claim for reimbursement from Business Entity A to Business Entity B, the question arises as to whether to focus on the governing law of the contract between them (A and B), or the governing law of the contract between the Data Subject and Business Entity A. In cases where each Business Entity, acting as either a SaaS provider or an IaaS provider, has a shared role for the purpose of building the same platform service, it may be permissible, depending on circumstances, to recognize the governing law of the basic agreement for that platform service (i.e., the governing law of the contract between the Data Subject and Business Entity A) as the law of the place of closest relationship.

Incidentally, in situations like ① or ④ in Table 2, since there is no pre-existing contractual relationship among parties, claims shall be governed by the law designated by general choice-of-law provisions on tort. Whether pursuant to the Rome II Regulation or AGRALJ, in determining the law of the place where the damage has occurred, the court may consider which jurisdiction has been most affected by the alleged harmful act, taking the circumstances of the case into account.

In summary, in cases where a tortious event occurs between parties who already have a contractual relationship, and the alleged tortious act is somehow related to their contract, there is room to apply the same law to their contractual and non-contractual issues. Given the technical structure of data transfer practices summarized in Chapter 2, respecting the party autonomy in the choice-of-law process on non-contractual issues through the escape clause in the Rome II Regulation or the AGRALJ would be consistent with the legitimate expectations of the parties involved. From the perspective that the party autonomy principle helps the court respect private ordering, it is also expected that the Business Entities recognize the significant impact of their terms and conditions on personal data management, including their privacy policies, and exercise prudence when drafting or revising them.

## 5. Conclusion

This paper has proposed a typological classification of cross-border data transfers as they occur in the actual economy, and has advanced a legal analysis focusing on the basis of liability — whether contractual, non-contractual, or concurrent — as well as on the presence of foreign elements, with a view to examining the appropriate framework for determining the applicable law. In the course of this analysis, this paper first elucidated the technical characteristics of cloud computing — notably the fragmentation of data and its international distribution in the context of secret sharing — and demonstrated that, when these characteristics are taken seriously, traditional private international law measures predicated on the identification of a physical location may give rise to significant difficulties in the choice of applicable law.

Proceeding from the analysis of relationship among actors and of the bases of liability set out in the typological framework developed herein, and focusing on the prevailing pattern in practice— the concurrence of contractual liability and tortious liability — this paper has established that the concept of private ordering shall be employed to select the applicable law by reference to the law governing the contract, and that this constitutes an appropriate measure in so far as it enhances the foreseeability of the parties involved.

Where, however, no contractual relationship exists — as in scenarios ① and ④ of Table 2 — such a measure is not readily available, as it would instead be necessary to identify the place of damage from the perspective of the most closely connected law. In these scenarios, two distinct contracts exist — one between the Data Subject and Business Entity A, and another between Business Entity A and Business Entity B — the latter being concluded in furtherance of the former, creating a contractual chain between the two. By analogy to the debate on global supply chains, one possible approach would be to align the governing law of the latter with that of the former — or, more precisely, given that the former is governed by a law selected through private ordering as described above, to extend the application of private ordering to the latter as well. Whether such an approach is viable, however, requires further and more cautious examination: it is not self-evident that SaaS providers invariably hold superior bargaining power or financial capacity over IaaS/PaaS providers, and whether the reasoning developed in the global supply chain context can be transposed by analogy to cross-border data transfers remains an open and contested question. This issue falls outside the scope of the present paper and is reserved for future inquiry.

Be that as it may, the analysis presented herein provides a contemporary choice-of-law framework tailored to the technical realities of cloud computing and AI, focused on the cases most representative of actual practice. The paper thereby contributes a concrete legal prescription for resolving civil disputes arising from personal data leak in the context of cross-border data transfers. It is in this contribution that the significance of this paper resides, and with this observation it is brought to a close.